\documentclass{aa}
\usepackage{graphicx}
\usepackage{epsfig}
\usepackage{amssymb}
\usepackage[english]{babel}

\title{Thermal stability of self-gravitating, optically thin accretion disks}
\author{G. Bertin \and G. Lodato}
\institute{Scuola Normale Superiore, Piazza dei Cavalieri 7, I-56126 Pisa, 
Italy}
\offprints{G. Lodato}
\date{Received / Accepted}
\begin{document}
\titlerunning{Thermal stability of self-gravitating accretion disks}
\maketitle

\begin{abstract}
In the dynamics of accretion disks, the presence of collective effects
associated with the self-gravity of the disk is expected to affect not only the
momentum transport, but also the relevant energy balance equations, which could
differ substantially from the non-self-gravitating case. Here we follow the 
picture that, when the disk is sufficiently cold, the stirring due to 
Jeans-related instabilities acts as a source of effective heating. The 
corresponding reformulation of the energy equations allows us to: ({\it i}) 
demonstrate how self-regulation can be established, so that the stability 
parameter $Q$ is maintained close to a threshold value, with weak dependence on
radius; ({\it ii}) rediscuss the opacity properties in the self-gravitating 
regime. In particular, we show that, if cooling is dominated by {\it
bremsstrahlung}, an optically thin stationary accretion solution is thermally 
stable, even in the non-advective case, provided the disk is self-gravitating. 
The details of the cooling function have little effect on the structure of such
accretion disk, which is in any case brought by self-gravity to self-regulate. 
This condition of self-gravitating accretion is expected to be appropriate for 
the outer regions of many disks of astrophysical interest. With the 
reformulation of the energy equations described in this paper we have also 
secured:  ({\it iii}) a starting point for the study of the emission properties
of self-gravitating accretion disks; ({\it iv}) a tool to analyze the structure
of the transition region, where the disk becomes self-gravitating.         
\end{abstract}

\section{Introduction}

A key step in the modeling of accretion disks is the detailed description of 
the balance between heating and cooling terms in the energy transport equations
(see Pringle \cite{pringle}). In this context, two hypotheses have been 
generally considered. The first states that the only source of heating in the 
disk is viscous dissipation, so that the available power is provided by the 
release of gravitational binding energy from the infalling matter. The second 
hypothesis is that cooling is essentially radiative. This latter hypothesis 
leads to different disk models, depending on whether the disk is optically 
thick (as in the ``standard'' model of Shakura \& Sunyaev \cite{shakura}) or 
optically thin (as in the model of Shapiro, Lightman \& Eardley \cite{shapiro};
hereafter SLE). One more recent development in the description of the 
energetics of accretion disks has been the recognition that radiative cooling 
may sometimes be insufficient. Under these circumstances, most of the energy 
remains stored in the disk rather than being radiated away, leading to the 
so-called {\it advection-dominated accretion flows} (ADAF; Narayan \& Yi 
\cite{yi}). This situation is described by a new equation of energy transport 
relevant to the radial direction, including an advection term (see Sect. 
\ref{sec:standard} below).

Once the general characteristics of stationary models are established, one can
proceed to address the problem of their stability. Different types of 
instabilities have been considered, leading to important conclusions about the 
possibility of applying the models to concrete systems. In one line of 
research, the stability analysis is carried out with the goal of identifying
microscopic collective mechanisms able to justify the viscosity
$\alpha$-prescription that is commonly introduced, following Shakura \& Sunyaev
(\cite{shakura}), in order to bypass the low level of viscosity predicted by
classical arguments. Other interesting stability aspects have also been 
investigated. In particular, the thermal-viscous instability related to  
hydrogen ionization (Meyer \& Meyer-Hofmeister \cite{meyer}) has been invoked 
to explain the variability of dwarf novae. In contrast, the optically thin SLE 
model has been commonly regarded as inapplicable, because of the thermal 
instability that is found to occur when optically thin {\it bremsstrahlung} is 
the dominant cooling mechanism. To be sure, the thermal stability properties of
the models depend strongly on the assumptions about the balance between heating
and cooling terms (for example, it is well known that, in the ADAF context, 
optically thin solutions exist in the inner disks and are thermally stable; see
Section \ref{sec:standard} and Narayan \& Yi \cite{yi2}). Indeed, it has been 
noted (Piran \cite{piran}) that small variations in the viscosity law can 
change the stability of accretion disk models drastically.

The self-gravity of the disk can change this picture significantly. It may have
an obvious impact on the radial and on the vertical force relations 
(Paczy\'nski \cite{pacz}). Furthermore, it has long been recognized that 
non-axisymmetric instabilities associated with the self-gravity of the disk can
affect the properties of angular momentum transport and hence the relevant 
viscosity law (e.g., see Lin \& Pringle \cite{lin2},  Laughlin \& Bodenheimer 
\cite{laugh}), although it still remains to be clarified whether such 
modifications can be reconciled with the standard $\alpha$-prescription.
Recently we have argued (Bertin \cite{bertin}, Bertin \& Lodato \cite{lodato})
that self-gravity should be properly incorporated also in the energy balance
equations, which should thus take a more general form than usually considered. 
In these first investigations we actually {\it replaced} the energy equations 
by a self-regulation prescription demanding that the axisymmetric stability 
parameter $Q$ is maintained close to a value $\bar{Q}$ basically dictated by 
the condition of marginal Jeans stability. The physical justification for the 
existence of 
an efficient dynamical thermostat of this kind is based on the presence of 
suitable competing heating and cooling mechanisms, but was not made explicit in
terms of a set of consistent equations for the energy transport. This 
simplified approach is adequate for the construction of dynamical models, and 
has already led to some unexpected results (in particular, the natural 
occurrence of flat rotation curves). Yet, it deliberately avoids the issue of
what is the predicted emission of the accretion disk, which has had such an
important role in the confrontation of models with observations in this general
research area. We anticipate that the disk self-gravity should be significant
only in the cold outer disk. In this respect, the identification of a 
consistent set of energy equations would also be desirable in order to set up
a proper matching of solutions across the relevant transition region.

In this paper, we include the effect of Jeans-related instabilities in the 
energy balance equations by means of a term of effective heating. Such heating
term is modeled as a function steeply rising when $Q$ falls below a threshold
value $\bar{Q}$ and practically vanishing when $Q$ rises above $\bar{Q}$.
We then show how this more general energy equation leads to the 
self-gravitating, self-regulated accretion disk solution, characterized by a 
value of $Q$ very close to $\bar{Q}$ and only weakly dependent on radius. This
stationary solution, although non-advective and optically thin, turns out to be
thermally stable.

The paper is organized as follows. In Section \ref{sec:standard} we briefly
summarize the arguments that lead to the balance of heating and cooling terms 
in current theories of accretion disks. In Section \ref{sec:effect} we 
propose a way to incorporate the effects of self-gravity in the energy 
equations by means of an additional heating term. In Section \ref{sec:equation}
we then provide and discuss a complete set of equations for the construction of
the full models of self-gravitating accretion disks. In Section \ref{sec:stab} 
we outline the structure of the stationary models and study their thermal
stability. The conclusions of the paper are given in Section \ref{sec:concl},
where we briefly comment on possible astrophysical applications.

\section{Energy balance in non-self-gravitating disks}
\label{sec:standard}

A generally accepted approach to the construction of accretion disk
models starts from 
the assumption that the disk is geometrically thin (so that vertically 
integrated and vertically averaged quantities can be usefully introduced) and
adopts a fluid model, specified by the continuity equation, the radial 
component of the Euler equation, the conservation of angular momentum in the 
presence of viscous torques, a viscosity prescription (such as the 
$\alpha$-prescription), an expression for the vertical thickness of the disk
(to be derived from the solution of the vertical hydrostatic equilibrium 
equations), and the relevant energy equations. In the following, we will refer 
to vertically integrated or averaged quantities, such as the surface density
$\sigma$, as ``surface'' quantities, while we will refer to other quantities, 
such as the volume density $\rho$, as ``volume'' quantities.

The energy equations are generally based on two hypotheses, of purely viscous
heating (with the addition, in some cases, of irradiation from the inner disk
or from the central object) and of purely radiative cooling. The first 
hypothesis is thus contained in an expression for the heating rate $Q^+$ 
in terms of ``surface quantities'':
\begin{equation}
\label{qpiu}
Q^+=\nu\sigma(r\Omega')^2\equiv D(r),
\end{equation}
where $\nu$ is the viscosity coefficient and $\Omega=\Omega(r)$ is the 
differential rotation (usually taken to be Keplerian).

The second hypothesis, that cooling is entirely radiative, implies that the
cooling rate per unit surface $Q^-$ can be calculated from the ``volume'' 
properties of the disk and the assumed radiation mechanisms:
\begin{equation}
\label{qmeno}
Q^-=R.
\end{equation}

The energy equations are then split into an equation for the radial transport
and an equation for the vertical transport. In practice, in systems with given
physical properties, one should carefully check the validity of this 
separation, which is natural only in the limit of thin disks. For the 
horizontal transport the equation is the so-called ``advection'' equation:
\begin{equation}
\label{adv}
\sigma u T_0\frac{\mbox{d} s}{\mbox{d} r}=Q^+-Q^-,
\end{equation}
where $u$ is the radial velocity, $T_0$ is the midplane temperature,
and $s$ is the entropy per unit mass. This 
equation states that, locally, the unbalance between energy gained from heating
and energy lost from cooling is stored in the disk and transported radially 
along with the infalling matter. Entropy in Eq. (\ref{adv}) is calculated based
on the first principle of thermodynamics:
\begin{equation}
\label{first}
T\mbox{d} s_v=\mbox{d} Q=c_v\mbox{d} T+p\mbox{d}(1/\rho),
\end{equation}
where $\rho$ is the volume density, $p$ is the pressure, $c_v$ is the 
specific heat at constant volume and $s_v$ is the ``volume'' entropy. After 
averaging over the vertical direction one defines the averaged ``surface''
entropy so that:
\begin{equation}
\label{entropy}
T_0\frac{\mbox{d} s}{\mbox{d} r}=\frac{1}{\gamma-1}\frac{\mbox{d} c^2}
{\mbox{d} r}-\frac{c^2}{\rho_0}\frac{\mbox{d}\rho_0}{\mbox{d} r},
\end{equation}
where $c$ is the thermal speed on the equatorial plane, $\rho_0$ is the 
central volume density, so that $\sigma=2\rho_0h$ ($h$ is the geometrical 
thickness) and $\gamma$ is the specific heat ratio. Note that in Eq. 
(\ref{adv}) one neglects the effects of horizontal radiative energy transport 
(but see Eq. (3) in Narayan \& Popham \cite{popham}). For the vertical 
transport one uses the equations of radiative transport, which, integrated over
the $z$-coordinate, give rise to Eq. (\ref{qmeno}).

A final remark is in order. The above equations are a set of differential 
equations, and hence they should be supplemented by a set of appropriate
boundary conditions. However, for standard disks (see Sect. 
\ref{sec:standard1} below), in which one neglects advection 
terms, the system reduces to a set of algebraic equations and no boundary 
conditions are thus required. In contrast, ADAF solutions are based on the full
differential problem. Even in this case, however, the self-similar solution 
found by Narayan \& Yi (\cite{yi}) (see Sect. \ref{sec:adaf}) avoids the need 
of specifying the boundary conditions. Other ADAF models take such 
self-similar solution as the approximate solution to be matched at
large radii. In this context, it has been noted (Yuan \cite{yuan}) that the 
boundary conditions are, indeed, important in determining the structure of the 
accretion disk and should be handled carefully.

\subsection{Standard disk}
\label{sec:standard1}
The so-called standard disk solutions are those for which, in Eq. (\ref{adv}), 
the dominant terms are $Q^+$ and $Q^-$, so that the advection equation reduces 
to:
\begin{equation}
\label{sei}
Q^+\approx Q^-,
\end{equation}
If we express this relation in view of the physical hypotheses mentioned 
previously, then the energy balance equation for standard disks is:
\begin{equation}
\label{standa}
D=R.
\end{equation}

As $R$ describes the emission of radiation from the disk, Eq. (\ref{standa})
shows that the emission is related to $D$ in a simple way, largely independent
of the detailed construction of the model. On the other hand, Eq. 
(\ref{standa}), involving all the surface quantities ($\sigma$, $T_0$,...), is
the closure relation needed to calculate the structure of the accretion model. 

\subsection{ADAF models}
\label{sec:adaf}
In some cases, the output of radiation may be insufficient so that the
balance between $Q^+$ and $Q^-$ in Eq. (\ref{adv}) is not possible. In this
more general case one cannot use Eq. (\ref{sei}) (and hence Eq. 
(\ref{standa})). This situation has often been described in terms of a 
convenient parameter, called $f$, defined as:
\begin{equation}
\label{f}
f=1-\frac{R}{D}
\end{equation}
so that Eq. (\ref{adv}) can be rewritten as:
\begin{equation}
\label{adve}
\sigma u T_0\frac{\mbox{d} s}{\mbox{d} r}=fD.
\end{equation}

For specified radiation processes, it is clear that Eq. (\ref{f})
and Eq. (\ref{adve}) together can close the model, which is characterized
(for given values of the viscosity parameter $\alpha$ and of the mass and 
angular
momentum accretion rates $\dot{M}$, $\dot{J}$) by a profile $f=f(r)$ determined
by Eq. (\ref{f}). In reality, it has been noted that, at a given radius $r$, by
considering the relation between $f$ and $\dot{M}$ ($\alpha$ and $\dot{J}$ 
being fixed), there are in general three branches of solutions (at least for 
sufficiently small values of $\dot{M}$): ({\it i}) one
branch corresponding to $f\ll 1$, so that (see Eq. (\ref{f})) $R\approx D$, 
is similar to the solution of standard disks, ({\it ii}) one new branch, called
ADAF, with $f\approx 1$, is characterized by the balance between viscous 
heating and the advection term (see Eq. (\ref{adve})); for this branch there is
little emission of radiation, as $R=(1-f)D\ll D$. Finally, a third branch 
({\it iii}), for intermediate values of $f$, is thermally unstable and 
corresponds to the SLE solution.

Treating $f$ as a given parameter, Narayan \& Yi (\cite{yi}) have shown that
models closed only by Eq. (\ref{adve}) (that is, not taking into account the
radiation properties) admit a simple, self-similar solution. This solution can
be seen as a first approximation to the ADAF solution on a wide (but, in any 
case, limited) radial range.

\section{Influence of self-gravity on the energy balance}
\label{sec:effect}

From the solution of the non-self-gravitating disk equations, it is interesting
to check the behavior of the parameter $Q$, which, as is well known, regulates 
the stability of the disk against axisymmetric Jeans instability. For a 
fluid one-component disk the parameter $Q$ is defined as:
\begin{equation}
\label{Q}
Q=\frac{c\kappa}{\pi G\sigma},
\end{equation}
where $\kappa$ is the epicyclic frequency (equal to $\Omega$ in the Keplerian 
case). Self-gravity becomes important at large radii. In fact, $Q(r)$ is 
expected to be a decreasing function of radius. For example, for the optically
thick disk dominated by the Kramers opacity (the outer region of the Shakura \&
Sunyaev \cite{shakura} disk) it is proportional to $r^{-9/8}$. When the 
standard model is applied to concrete cases (for example, the conditions 
typical of an AGN) $Q$ is found to plunge well below unity in regions of 
interest (Kumar \cite{kumar}). This behavior is untenable. It is only a 
reminder that the equations considered initially to construct the accretion
disk model must be reformulated so as to incorporate the important role of the
disk self-gravity (e.g., see Kolykhalov \& Sunyaev \cite{koly}).

As a quick, constructive way to address the problem of self-gravitating
accretion disks, we have recently argued (Bertin \cite{bertin}) that the energy
equations, when self-gravity is dominant, should be replaced by the physically
based prescription that $Q$ is self-regulated to a value $\bar{Q}$ close to 
unity. This is all we need if we are interested in calculating the main 
dynamical 
characteristics of such self-regulated accretion disks. On the other hand, a
detailed examination of the physical justification at the basis of the
self-regulation process (see Sect. 3 of the article by Bertin \& Lodato 
\cite{lodato}) readily shows that self-regulation results from heating and 
cooling mechanisms that go beyond those considered in Sect. \ref{sec:standard}.
In the self-regulation picture, widely considered in the study of spiral 
galaxies, the Jeans instability has the effect of inducing a fast heating of 
the disk (on the dynamical time-scale), at a rate that depends strongly on $Q$.
Such heating thus tends to bring $Q$ towards values of stability. On the other
hand, if the system is Jeans stable to begin with, efficient cooling mechanisms
(that, in some cases, may be just the radiative cooling of Eq. (\ref{qmeno})) 
can cool the disk down to lower values of $Q$. In the following Sections we 
will reformulate the energy balance equations starting from the arguments just 
presented. Such reformulation of the energy equations will allow us to: ({\it 
i}) demonstrate how self-regulation can be established; ({\it ii}) rediscuss
the opacity properties in the self-gravitating regime. In addition, it will
give us ({\it iii}) a starting point for the study of the emission
properties of self-gravitating accretion disks; ({\it iv}) a tool to
analyze the structure of the transition region, where the disk becomes
self-gravitating. A study of the last two points is postponed to a separate
investigation.

\subsection{Additional heating}

We suggest that in the expression of the heating rate a term $H_J$ should be 
added (so that $Q^+ = D + H_J$), strongly dependent on the value of
$Q$, which we express in terms of surface quantities as:
\begin{equation}
\label{accaj}
H_J=g(Q)\sigma c^2\Omega.
\end{equation}
Here $g(Q)$ is a function that rises quickly when $Q$ drops below $\bar{Q}$ 
(for example, $g(Q)\sim (\bar{Q}/Q)^n$, with $n\gg 1$). The form of this 
heating
term is argued on dimensional grounds, given the fact that the collective 
effects associated with self-gravity are expected to heat the disk on the 
dynamical time-scale. The general structure of $H_J$ is analogous to that of 
the heating term introduced in the context of spiral structure in galaxies to
demonstrate the time-dependent process of self-regulation (see Bertin \& Romeo 
\cite{romeo}). The argument is also analogous to that used by Lin \& Pringle 
(\cite{lin}) to incorporate the collective effects associated with self-gravity
into an effective viscosity. In fact, we could write (if we take $h=c^2/\pi G
\sigma$):
\begin{equation}
\label{eq:accavisc}
H_J=(\alpha_Jch)\sigma(r\Omega')^2,
\end{equation}
with the definition:
\begin{equation}
\label{alphaj}
\alpha_J=\frac{g(Q)}{Q}\frac{\Omega\kappa}{(r\Omega')^2}.
\end{equation}
With this device, we see that the term we are introducing mimics the viscous
heating for a system with modified viscosity $\alpha\rightarrow\alpha+
\alpha_J$. In practice, we think that the two issues, of a modified viscosity 
in the presence of spiral instabilities and of a heating term associated with
Jeans-related instabilities, should be kept separate. Note that in the context 
of spiral galaxies the emphasis is reversed, and little attention is paid to 
the problem of viscosity, although it is well known that 
spiral instabilities eventually favor mass accretion toward the center. In any
case, it should be stressed that the present approach remains heuristic. The
issue of energy dissipation requires further clarification (see also Balbus
\& Papaloizou \cite{balbus}).

\subsection{Additional cooling}
\label{sec:other}
By referring to the context of galactic disks, we find another important
element that is generally overlooked in the discussion of the energy balance
equations for accretion disks.

Basically this is the distinction between thermal speed and {\it effective} 
thermal speed. In practice, the effective thermal speed entering the dynamical 
equations may be higher than the thermal speed entering the radiation 
equations. The interstellar medium of galaxy disks is a cloudy medium, for
which the thermal speed associated with the temperature of the clouds is much
smaller than the velocity dispersion of the cold gas clouds. It is the latter
that sets the appropriate effective thermal speed relevant to the dynamics of
the disk. When a fluid model of the interstellar medium is adopted, in 
studies of disk dynamics, the numbers that are used for the equivalent acoustic
speed are then those applicable to the turbulent speed characteristic of the
cloudy medium. Therefore, one important cooling mechanism is that associated
with the inelastic collisions between gas clouds. Mechanisms of this kind may 
operate also in accretion disks, but they are usually ignored (but see
Quataert \& Chiang \cite{chiang}).

\subsection{Modified energy balance equation}

Because of these considerations we see that the energy balance equation for a
{\it non-advective} disk, should be written as:
\begin{equation}
\label{newheat}
D+H_J\geq R.
\end{equation}

In the inner disk $Q$ is likely to be much larger than unity: if we drop the
{\it caveat} about the distinction between thermal speed and effective thermal
speed, we see that such equation would then reduce back to that of a standard
disk.

\section{Basic equations for steady-state self-gravitating accretion disks
close to the transition region}
\label{sec:equation}

In the following we will study the problem of self-gravitating accretion
disks in a region not far from the transition region from non-self-gravitating
to self-gravitating disk (see Fig. \ref{fig:sketch}).

\subsection{Equations}

The analysis is thus simplified in two respects.

({\it i}) On the one hand, the radial balance of forces for a
cool, slowly accreting disk will be described by the Keplerian relation:
\begin{equation}
\label{kepler}
\Omega^2\simeq\frac{GM_{\star}}{r^3},
\end{equation}
where $M_{\star}$ is the mass of the central object. In Eq. (\ref{kepler}) we
have neglected the contribution of the disk self-gravity. This approximation is
valid for $r\ll r_s=2(GM_{\star})(\bar{Q}/4)^2(G\dot{M}/2\alpha)^{-2/3}$, where
$\alpha$ is the viscosity parameter (see Eq. (\ref{alpha}) below). For 
parameters typical of an AGN, previous studies (Bardou, Heyvearts \& Duschl 
\cite{bardou}) show that the transition from non-self-gravitating to 
self-gravitating disk occurs at a radius, which we call $r_Q$, that is much 
smaller than $r_s$, so that, at least for a certain parameter range, the 
approximation in Eq. (\ref{kepler}) is applicable to the inner parts of the 
disk. The effects associated with the integral relation between the 
gravitational field generated by the disk $\mbox{d}\Phi_{\sigma}/\mbox{d} r$ 
and $\sigma$ are described in the paper by Bertin \& Lodato (\cite{lodato}).

({\it ii}) On the other hand, the energy balance equation is taken to be (for 
the definitions of $D$ and $H_J$ see Eqs. (\ref{qpiu}) and (\ref{accaj})):
\begin{equation}
\label{balance}
D+H_J=R
\end{equation}
with $R$ specified by the relevant radiation mechanism. In the following we 
will use the expression for the cooling rate per unit volume appropriate for 
optically thin {\it bremsstrahlung} (but see the additional discussion of Sect.
\ref{robustness}):
\begin{equation}
\label{brems}
q_{br}=\sqrt{\frac{2\pi kT}{3m_e}}\frac{2^4 e^6}{3\hbar m_ec^3}\bar{g}n^2
\simeq1.4\, 10^{-27}T^{1/2}\bar{g}n^2,
\end{equation}
where $n$ is the particle density, given by $n=\rho/\mu m_p$ ($\mu$ is the mean
molecular weight), and $\bar{g}\approx 1.2$ is a gaunt factor averaged over
frequencies (see Rybicki \& Lightman \cite{lightman}), and where the last
expression is evaluated in CGS units ($T$ measured in K). In Eq. (\ref{brems})
$c$ is the speed of light and should not be confused with the effective thermal
speed defined above; we have assumed that the atomic weight is unity, and that 
$n_i=n_e$. In the spirit of the thin disk approximation we will define 
$R=q_{br}(z=0)h$. The optically thin description breaks down at small radii, 
below a scalelength $r_T$, where the disk is optically thick (see discussion in
Sect. \ref{sec:stab}).

The remaining equations are the equation of continuity:
\begin{equation}
\label{cont}
\dot{M}=-2\pi ru\sigma,
\end{equation}
the conservation of angular momentum:
\begin{equation}
\label{jpunto}
\dot{J}=\dot{M}r^2\Omega+2\pi\nu\sigma r^3\frac{\mbox{d}\Omega}{\mbox{d} r},
\end{equation}
and the standard $\alpha$-prescription (Shakura \& Sunyaev \cite{shakura}) for
viscosity:
\begin{equation}
\label{alpha}
\nu=\alpha c h,
\end{equation}
where $h$ is the vertical thickness of the disk, defined by:
\begin{equation}
\label{acca}
h=\frac{c^2}{\pi G\sigma}\frac{\pi}{4Q^2}\left[\sqrt{1+8Q^2/\pi}-1\right].
\end{equation}
The last equation is an interpolation formula for the ``exact'' thickness 
calculated including both the self-gravity of the disk and the gravitational 
force of the central object in the vertical hydrostatic equilibrium (see Eq. 
(A9) in the article by Bertin \& Lodato \cite{lodato} and discussion below),
in the limit of Keplerian rotation curve. Note that in the limit of
non-self-gravitating disk ($Q\gg 1$) this equation leads to the commonly used
result $h=\sqrt{\pi/2}(c/\Omega)$. In the following we will take $\dot{J}=0$; 
the effects of a non-zero net angular momentum flux modify the disk structure 
only in the innermost parts of the disk.

The level of viscosity present, as measured by the parameter $\alpha$, should
depend also on the development of spiral instabilities, and, therefore, on the
parameter $Q$. Since for the self-gravitating part of the disk $Q$ will turn 
out to be self-regulated, we prefer to simplify the analysis by leaving 
$\alpha$ as an assigned parameter. Our reference value (often considered for
AGN-type accretion disks) will be $\alpha=0.05$.

\subsection{Characterization of the transition region}

\begin{figure}[hbt!]
  \resizebox{\hsize}{!}{\includegraphics{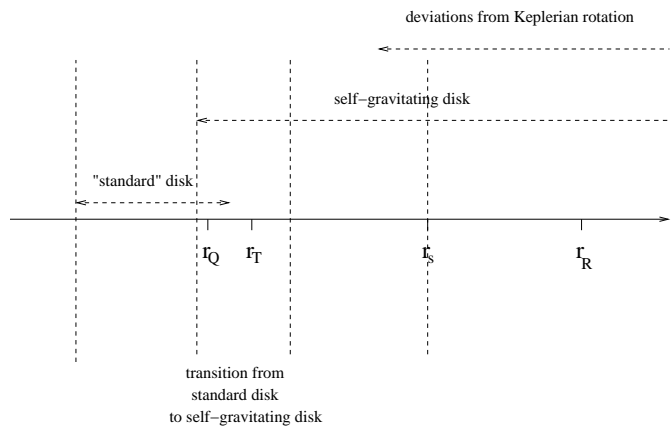}}
  \caption{\small{Sketch of the disk structure and of the different zones with 
the corresponding scale lengths. In practice, the present paper
  addresses the third zone, between $r_T$ and $r_s$.}}
  \label{fig:sketch}
\end{figure}

The above set of equations is naturally reduced to dimensionless form by the 
introduction of the length-scale:
\begin{equation}
\label{lengthscale}
r_R=\frac{4}{\mu m_pm_e}\left[\sqrt{\frac{\pi}{3}}\bar{g}\frac{e^6}{\hbar c^3}
\frac{4}{9\alpha}\left(\frac{G\dot{M}}{3\alpha}\right)^{-1/3}\frac{(GM_{\star})
^{1/2}}{2\pi G}\right]^{2/3}.
\end{equation}
In practice, this length-scale roughly corresponds to the outermost radius 
for which our solution exists (see Sect \ref{sec:branch} below). In the
following we will thus refer to the dimensionless radius $\hat{r}=r/r_R$.

Based on the scale lengths introduced so far, and on general results that will 
be better justified in Sect. \ref{sec:stab}, we can characterize the structure
of the accretion disk, in the region we are focusing on, by the following four
zones (see Fig. \ref{fig:sketch}): ({\it i}) An inner standard disk
with the value of $Q$ decreasing rapidly outwards; 
({\it ii}) A transition zone, in which the heating term due to self-gravity 
becomes important and the optical thickness of the disk decreases until 
({\it iii}) the optically thin regime is reached; ({\it iv}) At radii of the 
order of $r_s$ self-gravity affects also the rotation of the disk and at 
even larger radii the simple set of equations adopted in this Section breaks 
down and possibly other cooling mechanisms come into play. 

\subsection{Additional remarks on the role of self-gravity in the region
considered}

The term ``self-gravitating" is often used with different meanings, because
the quantity $\pi G \sigma$ may be judged to be important in relation to
different quantities: (a) The cumulative effect of $\pi G \sigma$ may make
the gravitational force in the plane different from the Keplerian field
generated by the central point mass; in this respect, to check whether the
disk should be considered to be self-gravitating, we should compare the
location $r$ of interest with the scale $r_s$ of the problem. In the limit
$r \gg r_s$ the gravitational field of the central point mass may be
neglected, but at $r\approx r_s$ the disk is already self-gravitating. (b) The 
transport properties in the disk may be affected significantly by density waves
and other Jeans-related phenomena; here the importance of the disk self-gravity
is quantified by comparing $\pi G\sigma$ to the acceleration $c \kappa$ and 
thus it is quantified by $Q$. Note that, in this respect, for a disk to be 
self-gravitating it is sufficient that $Q$ be close to unity; in other words, 
the self-gravitating limit is {\it not} set by the condition $Q \ll 1$, as 
might be thought at first. (c) The vertical scale of the disk may be 
significantly changed with respect to that of a non-gravitating disk; as shown 
in a separate article (Bertin \& Lodato \cite{lodato}; see Fig. A1 and Eq. (A9)
in the Appendix), this depends {\it both} on the local value of $Q$ {\it and} 
on the overall mass distribution in the disk. Furthermore, the expression of 
the disk thickness given in Eq. (A9) of Bertin \& Lodato (\cite{lodato}) is 
valid uniformly and can be considered as an ``exact" formula; the 
self-gravitating expression $h = c^2/\pi G \sigma$ is applicable, for example, 
when $Q\approx 1$ and the gravitational field in the disk plane is dominated by
the contribution of a disk characterized by $\sigma \sim 1/r$.

To a large extent, the three items above are independent aspects of the
role of self-gravity, even though we may argue that at very small radii any
accretion disk is likely to be non-self-gravitating from all the three
points of view, while a cold, radially extended accretion disk at large
radii is bound to be self-gravitating with respect to all aspects (a), (b),
and (c).

In this paper we are less interested in the specific impact of item (a)
(investigated earlier by us) and wish to focus on item (b), which had not
been properly explored in relation to the energy equations. Fortunately,
effects related to Jeans instability, which we model by means of the
heating term described by Eq. (\ref{accaj}), become important at a location
$r\approx r_Q$, which, for models applicable to AGN disks, occurs well far
in with respect to the location where item (a) comes into play, because
$r_Q \ll r_s$.  As to the disk thickness, we have just recalled that the
expression given in Eq. (A9) of Bertin \& Lodato (\cite{lodato}) can be
considered as an ``exact" formula; thus its Keplerian limit, as adopted in
Eq. (\ref{acca}), is legitimate when item (a) is unimportant. Therefore, in
the transition region we can safely follow the set of equations described
in subsection 4.1.

\section{Thermal stability of the self-regulated solution}
\label{sec:stab}

Optically thin, non-advective disks have been generally 
discarded because of the thermal instability that results from an increasing of
the heating rate relative to the cooling rate in response to an increase of the
temperature. This occurs because they are found to violate the condition for
thermal stability:
\begin{equation}
\label{stab1}
\frac{\mbox{d}}{\mbox{d} T}\left(\mbox{Log}\left(\frac{Q^-}{Q^+}\right)\right)
_{\sigma}>0,
\end{equation}
where the subscript $\sigma$ indicates that the derivative should be taken at
constant surface density (see, for example, Pringle \cite{pringle}).

Since our discussion of Sect. \ref{sec:effect} modifies the basic ingredients
of the energy balance equations, we should now check what are the properties
of the self-gravitating solutions in relation to the thermal stability
condition. Given the fact that the additional heating term $H_J$ given in Eq.
(\ref{accaj}) is a function of $Q$ and that such parameter is 
proportional to the thermal speed $c$, we can reformulate the thermal stability
condition as:
\begin{equation}
\label{stab2}
\frac{\mbox{d}}{\mbox{d} Q}\left(\mbox{Log}\left(\frac{Q^-}{Q^+}\right)\right)
_{\sigma}>0
\end{equation}

\begin{figure}[hbt!]
  \resizebox{\hsize}{!}{\includegraphics{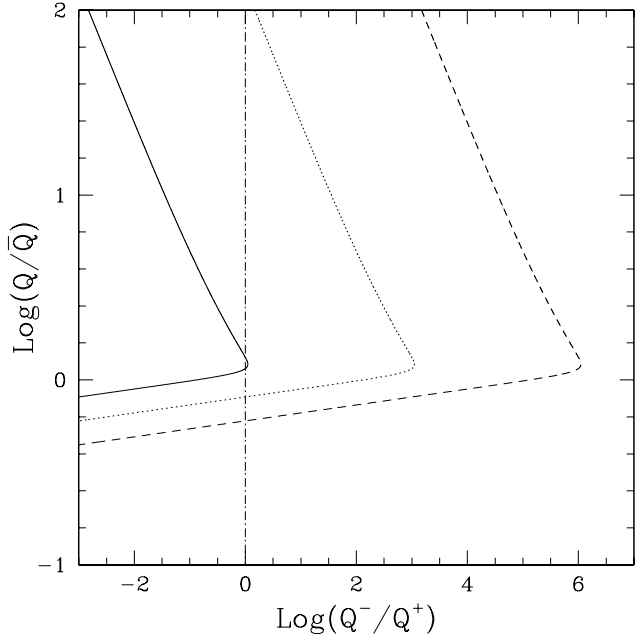}}
  \caption{\small{Relation between the ratio of heating to cooling rates and
the stability parameter $Q$ at three different radii, i.e., from 
right to left $\hat{r}=10^{-4}, 10^{-2}, 1$. The stable branch is the one 
with positive slope. The intersection between the vertical line and the stable 
branch is the stable self-regulated solution.}}
  \label{fig:stab}
\end{figure}

\subsection{The thermally stable self-gravitating branch}
\label{sec:branch}
In Fig. \ref{fig:stab} we show the relation between $Q^-/Q^+$ and $Q$ at fixed
radius, for different values of the radius. Equilibrium corresponds to  
$\mbox{Log}(Q^-/Q^+)=0$ (the vertical line in Fig. \ref{fig:stab}) and the 
solution is thermally stable if the slope of the curve is positive. As is 
easily recognized, there are two branches of solution: one, unstable, is 
characterized by $Q\gg 1$, for which the dominant heating mechanism is viscous 
dissipation as in non-self-gravitating disks (this solution can be regarded as 
analogous to the SLE solution). The other solution is self-gravitating (it has 
$Q\approx 1$) and it is, indeed, thermally stable. Note that such stable 
equilibrium solution is always characterized by a value of $Q$ close to unity. 
The equilibrium $Q$ increases slightly with radius (although this dependence is
weak); independently of the physical parameters of the disk, the maximum value 
of $Q$ for the stable solution is $\approx 1.3\bar{Q}$. There is a maximum 
radius for which this kind of solution is possible, located near $r_R$. The 
relative location of the two scale lengths $r_R$ and $r_s$ defines the 
importance of the deviation from Keplerian rotation in our model. For 
parameters typical of AGN configurations ($M_{\star}=10^8 M_{\sun}, 
\dot{M}=1M_{\sun}/yr\approx 6\, 10^{25}g/s$) we estimate $r_R/r_s\approx 5\, 
10^2$. Hence, before the scale $r_R$ is reached the rotation curve is expected 
to depart from the simple Keplerian curve, so that the quantitative results of 
the present simplified work are only approximate when $\hat{r}>2\, 10^{-3}$ and
may even be qualitatively incorrect when $\hat{r}$ is of order of unity. The 
lack of solutions at large radii ($\hat{r}>1$) is due to the fact that 
radiative cooling is insufficient and unable to balance the heating rate. In 
this context, one should keep in mind that, as we indicated in Sect. 
\ref{sec:other}, other kinds of effective cooling could be efficient and play 
a role in the outer regions, such as inelastic collisions between clouds, which
have been neglected here.

\subsection{Consistency of the optically thin description}

We have also calculated the optical depth of the disk, to check that the model
of optically thin disk is consistent. We express the ``effective'' optical 
depth $\tau^{*}$ (Rybicki \& Lightman \cite{lightman}) for the case 
$\kappa_{T}\gg \kappa_{br}$ (where $\kappa_{T}$ and $\kappa_{br}$ are the 
opacity due to Thomson scattering and {\it bremsstrahlung}, respectively), as:
\begin{equation}
\label{tau}
\tau^{*}=\sqrt{\tau_{br}\tau_{T}},
\end{equation}
where $\tau_{br}=\kappa_{br}\sigma$ and $\tau_{T}=\kappa_{T}\sigma$.

\begin{figure}[hbt!]
  \resizebox{\hsize}{!}{\includegraphics{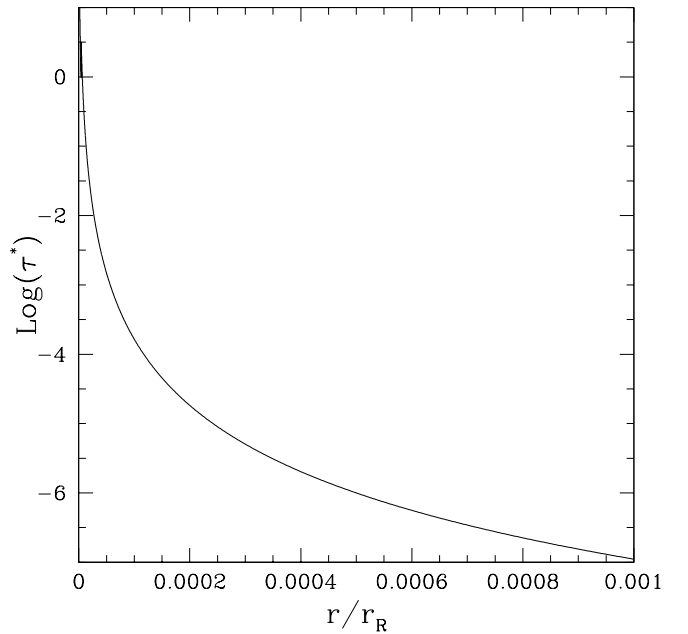}}
  \caption{\small{Optical thickness of the disk as a function of radius for 
$\dot{M}=1M_{\sun}/yr$. In the innermost regions the disk is optically
thick.}}
  \label{fig:thickness}
\end{figure}
\begin{figure}[hbt!]
  \resizebox{\hsize}{!}{\includegraphics{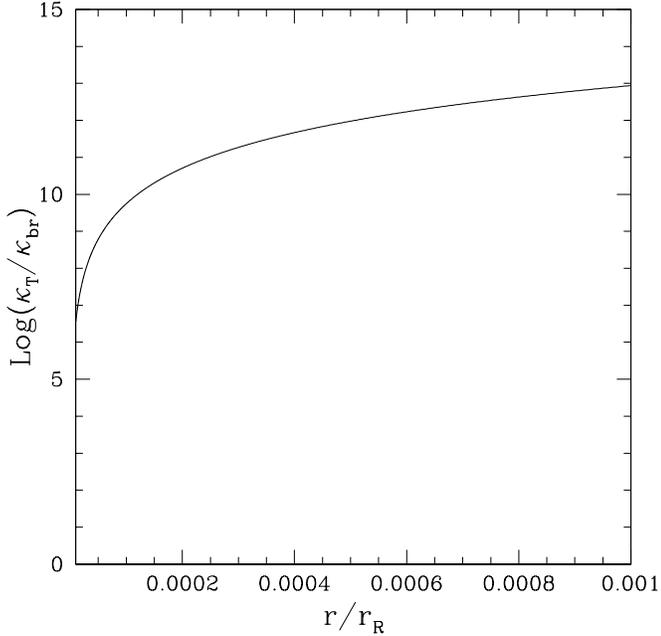}}
  \caption{\small{Ratio of Thomson opacity to free-free opacity, showing that
Thomson scattering always dominates over {\it bremsstrahlung}. }}
  \label{fig:thickness2}
\end{figure}

In Fig. \ref{fig:thickness} we show the optical depth as a function of radius 
for the thermally stable equilibrium solution. The disk is optically 
thin in the outer regions and the optical depth increases inwards. In the 
innermost regions the disk eventually becomes optically thick. This occurs at a
dimensionless radius $\hat{r}_T$ independent of the central mass and
very weakly dependent on the mass accretion rate (approximately 
$\hat{r}_T\propto\dot{M}^{-1/18}$); for $\dot{M}=1M_{\sun}/yr$
the transition to the optically thick region occurs at an inner radius 
$\hat{r}_T=r_T/r_R\approx 4\, 10^{-6}$. In Fig. \ref{fig:thickness2} we show
the profile of the ratio $\kappa_T/\kappa_{br}$ to check that Thomson opacity 
always dominates over free-free absorption. 

In the transition region from optically thick to optically thin emission, for a
proper matching, a more detailed description would be desired.

\subsection{Robustness with respect to changes of the radiative processes}
\label{robustness}
In contrast with other cases for which optically thin emission is considered 
(i.e., the inner regions of disks with very low mass accretion rates), in our 
case the temperature of the disk (for reasonable values of $\dot{M}$)
is expected to be low, so that pure
{\it bremsstrahlung} fails to be the dominant radiative process and line
emission and absorption become important. In view of this remark, we have
checked the consequences of a different cooling function, using a functional 
interpolation (Sarazin \& White \cite{sarazin}) of numerical data (Raymond, Cox
\& Smith \cite{raymond}), appropriate for lower temperatures. 

\begin{figure}[hbt!]
  \resizebox{\hsize}{!}{\includegraphics{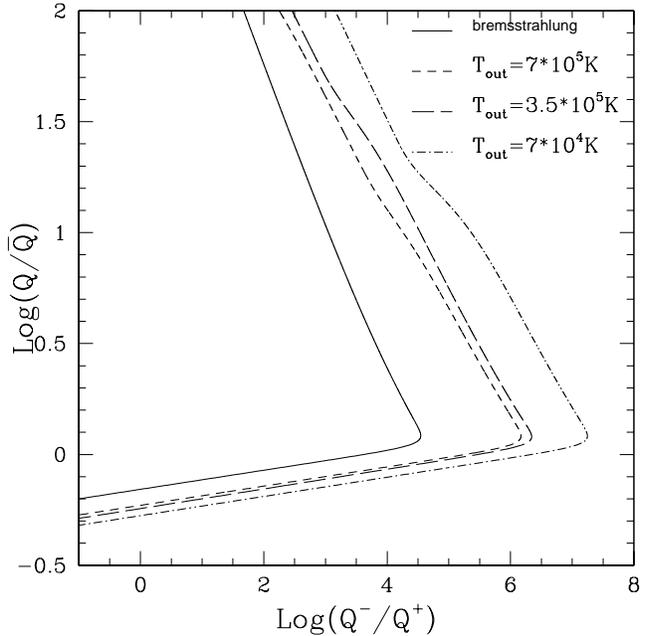}}
  \caption{\small{Stability curves for different choices of the cooling 
function at radius $\hat{r}=10^{-3}$. The details of the cooling
mechanism do not change the main characteristics of
on the stable self-gravitating branch of solutions. The three outer 
temperatures correspond to $\dot{M}\simeq 90,~30, ~3 M_{\sun}/yr$,
respectively. Here the value $\alpha=0.05$ has been assumed.}}
  \label{fig:stabcool}
\end{figure}

In this test of robustness, the value of the accretion rate $\dot{M}$, through
the temperature of the outer disk defined as $kT_{out}=(\mu m_p/2) (G\dot{M}/3
\alpha)^{2/3}$, appears explicitly in the problem. This is the result of the 
fact that the various line emissions become efficient at specified 
temperatures. In practice, only minor differences are found for the 
self-gravitating branch (even under extreme conditions; see
Fig. \ref{fig:stabcool}), with respect to
the previous analysis, because the cooling function influences 
only the equilibrium value of $Q$ but this is, in any case, self-regulated to
values close to unity (on the thermally stable equilibrium solution).

\section{Discussion and conclusions}
\label{sec:concl}

Self-gravitating accretion disks may be present in many astrophysical systems,
ranging from protostellar disks to AGN and could also be of interest for the 
general galactic context.

In protostellar disks it is likely that, especially at the 
beginning of the accretion processes, during which the mass of the central 
object is small, self-gravity plays an important role. The influence of 
self-gravity has been sometimes invoked to explain the flat infrared 
spectrum of some T Tau stars (Adams, Lada \& Shu \cite{shu}; Duschl, 
Strittmatter \& Biermann \cite{duschl}). However, current estimates for the
mass accretion rates in these systems (typically $\dot{M}\sim 10^{-8}M_{\sun}
/yr$) so far have discouraged interpretations of the shape of the
spectrum in terms of self-gravitating, optically-thick disks.

On the other hand, observational evidence suggests that, in some cases, 
the currently accepted models may be inadequate to describe the accretion flow 
of some AGN. For example, in the case of the Seyfert galaxy \object{NGC1068}, 
deviations from Keplerian rotation have been observed; when the standard 
Shakura \& Sunyaev (\cite{shakura}) model is applied to this case, the values 
of the stability parameter $Q$ in the outer disk turn out to be exceedingly low
(Kumar \cite{kumar}). In fact, the energy balance in standard disks, for which 
viscous dissipation is exactly balanced by radiative heating, leads to a $Q$ 
profile rapidly decreasing outwards. 

In this paper, we have focused on the problem of the energy budget in 
self-gravitating accretion disks, with the ultimate goal of providing a 
framework to describe the emission and the spectrum associated with such disks.
We have adopted the view that the effects of self-gravity on the energy balance
are not only related to viscous dissipation, but that there are also processes,
such as the self-regulation of Jeans instability, that should give a 
significant contribution to the heating of the disk, at least in its outer 
parts.

By referring to conditions applicable to some AGN, we have then shown that a 
thermally stable, non-advective, optically thin equilibrium solution is 
possible in the region dominated by self-gravity. In contrast with other 
optically thin solutions (Shapiro, Lightman \& Eardley \cite{shapiro}), our 
solution is thermally stable as a result of the strong dependence of the 
heating term related to self-gravity on the parameter $Q$. We have also found 
an unstable solution, which is not self-gravitating  ($Q\gg 1$), and we argue 
that it is a modification of the SLE solution.

The value of $Q$ on the thermally stable solution is always close to unity, and
it is slightly dependent on radius. The temperature in the outer disk is 
constant and depends on the mass accretion rate $\dot{M}$. The structure of the
disk, as long as the energy balance is dominated by self-gravity, turns out to 
depend only weakly on the detailed radiation processes involved. This has been
shown by taking into account two different radiation processes: pure {\it 
bremsstrahlung} and the case of a cooling function dominated by line absorption
and emission (Sarazin \& White \cite{sarazin}), which is best suited for the 
low temperature conditions expected in the outer disk. The dependence on the 
radiation processes is only weak because of the self-regulation mechanism: 
whatever the radiation process, the disk readjusts, so as to keep $Q$ close to
marginal stability.

For simplicity, in this paper we have focused on conditions relevant to the 
region of the disk where the transition to the self-gravitating regime occurs.
Further out, one should include the presence of additional cooling mechanisms,
not considered here, and especially the effects related to the deviations from
a Keplerian rotation curve. The latter effects have been widely discussed in a 
previous paper (Bertin \& Lodato \cite{lodato}).

\end{document}